\documentstyle[prd,aps,epsfig]{revtex}

\newcommand{\be}{\begin{equation}}
\newcommand{\ee}{\end{equation}}
\newcommand{\ba}{\begin{eqnarray}}
\newcommand{\ea}{\end{eqnarray}}
\newcommand{\ban}{\begin{eqnarray*}}
\newcommand{\ean}{\end{eqnarray*}}

\begin{document}

\title{CHROMO-HYDRODYNAMIC APPROACH TO THE UNSTABLE QUARK-GLUON PLASMA}

\author{Cristina Manuel\footnote{Electronic address:
{\tt cmanuel@ieec.uab.es}}}

\address{\it Institut d'Estudis Espacials de Catalunya (IEEC) \\
and Instituto de Ciencias del Espacio (CSIC) \\
Campus Universitat Aut\` onoma de Barcelona \\
Facultat de Ci\` encies, Torre C5 \\
E-08193 Bellaterra (Barcelona), Spain}

\author{Stanis\l aw Mr\' owczy\' nski\footnote{Electronic address:
{\tt mrow@fuw.edu.pl}}}

\address{\it So\l tan Institute for Nuclear Studies \\
ul. Ho\.za 69, PL - 00-681 Warsaw, Poland \\
and Institute of Physics, \'Swi\c etokrzyska Academy \\
ul. \'Swi\c etokrzyska 15, PL - 25-406 Kielce, Poland}

\date{5-th October 2006}

\maketitle

\begin{abstract}

We derive hydrodynamic-like equations that are applicable to
short-time scale color phenomena in the quark-gluon plasma.
The equations are solved in the linear response approximation, 
and the gluon polarization tensor is derived. As an application, 
we study the collective modes in a two-stream system and find 
plasma instabilities when the fluid velocity is larger than the
speed of sound in the plasma. The chromo-hydrodynamic approach, 
discussed here in detail, should be considered as simpler over 
other approaches and well-designed for numerical studies of the 
dynamics of an unstable quark-gluon plasma.

\end{abstract}




\section{Introduction}


Bulk features of electromagnetic plasmas are usually studied by
means of  fluid equations \cite{Kra73}. To get more detailed
information, one refers to kinetic theory \cite{Kra73}. Since the
fluid equations are noticeably simpler than the kinetic ones, the
hydrodynamic approach is also frequently used in numerical simulations
of the plasma evolution, studies of the nonlinear dynamics, etc.
\cite{Kra73}. The situation in the field of quark-gluon plasma studies 
is rather different. Although chromo-hydrodynamics equations were discussed
by several authors over a long period of time \cite{Kajantie:1980jj,Holm:hg,Holm:yh,Mrowczynski:1987ch,Mrowczynski:1989bv,Bhatt:1988tq,Jackiw:2000cd,Bistrovic:2002jx,Manuel:2003zr,Dai:2006te,Bambah:2006yg},
they were not carefully studied. The field of their applicability
was not established and very few important results were obtained by
means of them. Consequently, the chromo-hydrodynamics has not attracted
much attention. Instead, field theory diagrammatic methods have been
successfully applied to reveal equilibrium properties of the quark-gluon
plasma \cite{Blaizot:2001nr}, while  transport theory
\cite{Elze:1989un,Mrowczynski:np,Litim:2001db} proved efficient in 
the non-equilibrium studies. In particular, the important role of 
instabilities in the quark-gluon plasma evolution was clarified within 
the kinetic theory approach, see the review \cite{Mrowczynski:2005ki}. 
The kinetic equations were also a basis of extensive numerical simulations 
of the unstable QCD plasma 
\cite{Rebhan:2004ur,Dumitru:2005gp,Arnold:2005vb,Rebhan:2005re,Arnold:2005ef,Dumitru:2006pz,Romatschke:2006wg}.
The early stage of relativistic heavy-ion collisions, when the quark-gluon
system is produced, was effectively studied using methods of  classical
field theory, see {\it e.g.} the review  \cite{Iancu:2003xm} and very recent 
publications \cite{Romatschke:2005pm,Lappi:2006fp}.

Inspired by the success of hydrodynamic methods in the electromagnetic
plasma, we discuss the approach to be applied to the quark-gluon plasma.
Before going to the main subject of our study, however, a very important
point has to be clarified.  Real hydrodynamics deals with systems
which are in local equilibrium, and thus it is only applicable at
sufficiently long time scales. The continuity and the Euler or
Navier-Stokes equations are supplemented by the equation of state
to form a complete set of equations. The equations can be derived
from  kinetic theory, using the distribution function of local
equilibrium, which by definition maximizes the entropy density,
and thus the function cancels the collision terms of the transport
equations. Such a real chromo-hydrodynamics was derived in
\cite{Manuel:2003zr} where the state of local equilibrium was
found, using the collision term of the Waldman-Snider form. The
chromo-hydrodynamics has occurred trivial in the sense that although
the local equilibrium can be colorful, all color components of
the plasma move with the same hydrodynamic velocity. Therefore,
chromodynamic effects disappear entirely once the system is
neutralized. It actually happens even before the local equilibrium
is achieved \cite{Manuel:2004gk}. Thus, there is no QCD analog of
the magneto-hydrodynamics which is well known in the electromagnetic
plasma. The magneto-hydrodynamic regime appears due to a large
difference of electron and ion masses which effectively slows
down a mutual equilibration of electrons and ions. Therefore,
at a relatively long time scale one deals with the charged electron
fluid in local equilibrium which moves in a passive background of
positive ions.

Since the hydrodynamic equations express the macroscopic conservation
laws, the equations hold not only for systems in local equilibrium
but for systems out of equilibrium as well. In particular, the equations
can be applied at time scales significantly shorter than that of local
equilibration. At such a short time scale the collision terms of
the transport equations can be neglected. However, extra assumptions
are then needed to close the set of equations, as the (equilibrium)
equation of state cannot be used. In the electromagnetic plasma physics,
several methods to close the system of equations were worked out.

In this paper we discuss not a real hydrodynamics describing the
quark-gluon plasma in local equilibrium, but the fluid equations
which are valid at much shorter time scale. The approach is designed
to study temporal evolution of the unstable QCD plasma. In
Sec.~\ref{sec-eqs} the fluid equations are derived from the kinetic
theory which is presented in Sec.~\ref{sec-kin}. The equations are
solved in the linear response approximation in Sec.~\ref{sec-res}
and the polarization tensor is derived. As an example, the collective
plasma modes in the two-stream system are analyzed in
Sec.~\ref{sec-2stream}. The paper is closed with a few remarks on
the hydrodynamic approach.

Throughout the paper we use the natural units with $c=\hbar=k_B =1$
and the metric $(1,-1,-1,-1)$.


\section{Kinetic theory}
\label{sec-kin}


In this section we briefly present the transport theory of quarks and
gluons \cite{Elze:1989un,Mrowczynski:np} which is used to derive
the fluid equations.

The distribution function of quarks $Q(p,x)$, where $p$ is the
four-momentum and $x$ is the four-position, is a hermitian
$N_c\times N_c$ matrix in color space (for a SU($N_c$) color group).
The distribution function is gauge dependent and it transforms under
a local gauge transformation $U$ as
\be
\label{Q-transform}
Q(p,x) \rightarrow U(x) \, Q(p,x) \, U^{\dag }(x) \; .
\ee
The distribution function of antiquarks, which we denote by $\bar Q(p,x)$,
is also a hermitian $N_c\times N_c$ matrix in color space it also
transforms according to Eq.~(\ref{Q-transform}). The distribution
function of gluons is a hermitian $(N_c^2-1)\times (N_c^2-1)$ matrix,
which transforms as
\be
\label{G-transform}
G(p,x) \rightarrow {\cal U}(x) \: G(p,x) \:{\cal U}^{\dag }(x) \;,
\ee
where
\be
{\cal U}_{ab}(x) = 2{\rm Tr}\bigr[\tau^a U(x) \tau^b U^{\dag }(x)] \ ,
\ee
with $\tau^a ,\; a = 1,...,N_c^2-1$ being the SU($N_c$) group generators
in the fundamental representation with ${\rm Tr} (\tau^a \tau^b) = \frac12
\delta^{ab}$.

The four-momentum $p$, which is the argument of the distribution
functions, is not constrained by the mass-shell condition but it
is taken into account in the momentum integration measure. We use
the notation
\be
\label{measure}
\int_p \cdots \equiv \int \frac{d^4 p}{(2\pi )^3} \:
2 \Theta(p_0) \delta (p^2) \;,
\ee
and the color current is expressed in the fundamental representation as
\be
\label{col-current}
j^{\mu }(x) = -\frac{g}{2} \int_p p^\mu \;
\Big[ Q( p,x) - \bar Q ( p,x)
- {1 \over N_c}{\rm Tr}\big[Q( p,x) - \bar Q ( p,x)\big]
+  2 \tau^a {\rm Tr}\big[T^a G(p,x) \big]\Big] \; ,
\ee
where $g$ is the QCD coupling constant. A sum over helicities,
two per particle, and over quark flavors $N_f$ is understood in
Eq.~(\ref{col-current}), even though it is not explicitly written down.
The SU($N_c$) generators in the adjoint representation are expressed
through the structure constants  $T^a_{bc} = -i f_{abc}$, and are
normalized as ${\rm Tr}[T^aT^b]= N_c \delta^{ab}$. The current can be
decomposed as $j^\mu (x) = j^\mu_a (x) \tau^a$ with
$j^\mu_a (x) = 2 {\rm Tr} (\tau_a j^\mu (x))$.

The distribution functions of quarks, antiquarks and gluons satisfy
the transport equations:
\ba
p^{\mu} D_{\mu}Q(p,x) + {g \over 2}\: p^{\mu}
\left\{ F_{\mu \nu}(x), \partial^\nu_p Q(p,x) \right\}
&=& C[Q,\bar Q,G] \;,
\label{transport-q}  \\ [2mm]
p^{\mu} D_{\mu}\bar Q(p,x) - {g \over 2} \: p^{\mu}
\left\{ F_{\mu \nu}(x), \partial^\nu_p \bar Q(p,x)\right\}
&=& \bar C[Q,\bar Q,G]\;,
\label{transport-barq} \\ [2mm]
p^{\mu} {\cal D}_{\mu}G(p,x) + {g \over 2} \: p^{\mu}
\left\{ {\cal F}_{\mu \nu}(x), \partial^\nu_p G(p,x) \right\}
&=& C_g[Q,\bar Q,G]\;,
\label{transport-gluon}
\ea
where $\{...,...\}$ denotes the anticommutator and $\partial^\nu_p$
the four-momentum derivative; the covariant derivatives $D_{\mu}$ and
${\cal D}_{\mu}$ act as
$$
D_{\mu} = \partial_{\mu} - ig[A_{\mu}(x),...\; ]\;,\;\;\;\;\;\;\;
{\cal D}_{\mu} = \partial_{\mu} - ig[{\cal A}_{\mu}(x),...\;]\;,
$$
with $A_{\mu }$ and ${\cal A}_{\mu }$ being four-potentials
in the fundamental and adjoint representations, respectively:
$$
A^{\mu }(x) = A^{\mu }_a (x) \tau^a \;,\;\;\;\;\;
{\cal A}^{\mu }(x) = T^a A^{\mu }_a (x) \; .
$$
The strength tensor in the fundamental representation is
$F_{\mu\nu}=\partial_{\mu}A_{\nu} - \partial_{\nu}A_{\mu}
-ig [A_{\mu},A_{\nu}]$, while  ${\cal F}_{\mu \nu}$ denotes the field
strength tensor in the adjoint representation. $C, \bar C$ and $C_g$
represent the collision terms which are neglected in our further
considerations, as we are interested in short-time-scale phenomena
which are `faster' than the collisions.

The transport equations are supplemented by the Yang-Mills equation
describing generation of the gauge field
\be
\label{yang-mills}
D_{\mu} F^{\mu \nu}(x) = j^{\nu}(x)\; ,
\ee
where the color current is given by Eq.~(\ref{col-current}).


\section{Fluid equations}
\label{sec-eqs}


We assume here that there are several streams in the plasma system, 
and then the distribution function of, say, quarks $Q(p,x)$ can be 
decomposed as $Q(p,x)=\sum_\alpha Q_\alpha(p,x)$ where the index 
$\alpha$ labels the streams. Since we are interested in the short-time
scale phenomena - `faster' than collisions - the complete quark
distribution function $Q(p,x)$ as well as the distribution function 
of every stream $Q_\alpha(p,x)$ satisfy the collisionless transport 
equation. The streams interact with each other only via the mean-filed 
which is generated in a self-consistent way by the current where every 
stream contributes. We note that the decomposition of the distribution 
function $Q(p,x)$ into streams is, in principle, not unique. However, 
if the stream distribution functions $Q_\alpha(p,x)$ solve the transport 
equations combined with the field generation equation, the complete 
distribution function solves the transport equation and the field 
generation equation as well.
  
Further analysis is limited to quarks but inclusion of anti-quarks
and gluons is straightforward. Integrating the collisionless transport 
equation (\ref{transport-q}) satisfied by $Q_\alpha$ over the four-momentum 
with the integration measure (\ref{measure}), one finds the covariant 
continuity equation
\be
\label{cont-eq}
D_\mu n^\mu_\alpha = 0 \;,
\ee
where $n^\mu_\alpha$ is $N_c\times N_c$ matrix defined as
\be
\label{flow}
n^\mu_\alpha (x) \equiv \int_p p^\mu Q_\alpha(p,x) \;.
\ee
The four-flow $n^\mu_\alpha$ transforms under gauge transformations
as the quark distribution function, {\it i.e.} according to
Eq.~(\ref{Q-transform}).

Multiplying the transport equation (\ref{transport-q}) by the four-momentum
and integrating the product with the measure (\ref{measure}), we get
\be
\label{en-mom-eq}
D_\mu T^{\mu \nu}_\alpha
- {g \over 2}\{F_{\mu}^{\;\; \nu}, n^\mu_\alpha \}= 0 \;,
\ee
where the energy-momentum tensor is
\be
\label{en-mom}
T^{\mu \nu}_\alpha (x) \equiv \int_p p^\mu p^\nu  Q_\alpha(p,x) \;.
\ee

We further postulate that $n^\mu_\alpha$ and $T^{\mu \nu}_\alpha$ 
have the form of an ideal fluid {\it i.e.}
\ba
\label{flow-id}
n^\mu_\alpha(x)  &=& n_\alpha (x) \, u_\alpha^\mu(x) \;,
\\[1mm]
\label{en-mom-id}
T^{\mu \nu}_\alpha(x)  &=& {1 \over 2}
\big(\epsilon_\alpha (x) + p_\alpha (x)\big)
\big\{u^\mu_\alpha (x), u^\nu_\alpha (x) \big\}
- p_\alpha (x) \, g^{\mu \nu}\;,
\ea
where the hydrodynamic velocity $u^\mu_\alpha$ is, as $n_\alpha$, 
$\epsilon_\alpha$ and $p_\alpha$, a $N_c\times N_c$ matrix. The 
anticommutator of $u^\mu_\alpha$ and $u^\nu_\alpha$ is
present in Eq.~(\ref{en-mom-id}) to guarantee the symmetry
of $T^{\mu \nu}_\alpha$ with respect to $\mu \leftrightarrow \nu$
which is evident in the definition (\ref{en-mom}). 

Since Eqs.~(\ref{flow-id}, \ref{en-mom-id}) are of crucial importance
for our further considerations, let us discuss them in more detail. 
There are two aspects of the ideal-fluid form of $n^\mu_\alpha$ and 
$T^{\mu \nu}_\alpha$ to be justified: the Lorentz structure and the 
matrix structure. We start the discussion with the Lorentz structure. 
Eqs.~(\ref{flow-id}, \ref{en-mom-id}) assume that $n^\mu_\alpha$ and
$T^{\mu \nu}_\alpha$ can be expressed as products of the Lorentz scalars 
$n_\alpha$, $\epsilon_\alpha$ and $p_\alpha$ and the four-vector 
$u^\mu_\alpha$. Such a Lorentz structure appears when the distribution 
function $Q_\alpha(p,x)$, which gives $n^\mu_\alpha$ and  
$T^{\mu \nu}_\alpha$ through Eqs.~(\ref{flow}, \ref{en-mom}), is isotropic
in the local rest frame where $u^\mu_\alpha =(1,0,0,0)$. One does not need 
to assume that the function is of local equilibrium form, the isotropy is 
a sufficient condition. And we note that the isotropy of the momentum 
distribution, required Eqs.~(\ref{flow-id}, \ref{en-mom-id}),  is not a
serious limitation because we actually consider not a single stream 
but a superposition of several streams. And then, one can easily model
an arbitrary momentum distribution as a sum of several momentum distributions,
everyone isotropic in its Lorentz rest frame. Thus, the Lorentz structure 
expressed by Eqs.~(\ref{flow-id}, \ref{en-mom-id}) seem to be well justified.

The problem of the matrix structure of Eqs.~(14, 15) is more complicated.
We first note that $n^\mu_\alpha$ and $T^{\mu \nu}_\alpha$ have to be 
matrix value functions as the distribution function $Q_\alpha(p,x)$ which 
defines them through Eqs.~(\ref{flow}, \ref{en-mom}) is a matrix in a color
space. The question is whether $n_\alpha$, $\epsilon_\alpha$, $p_\alpha$ 
and $u^\mu_\alpha$ have to be all matrix quantities. When the system under 
consideration is colorless, these quantities can be treated as scalars in 
the color space. However, as explicitly shown in Sec.~\ref{sec-res}, even 
small perturbations of these quantities induced by a color field are of 
matrix value. Therefore, $n_\alpha$, $\epsilon_\alpha$, $p_\alpha$ and $u^\mu_\alpha$ have to be all treated as matrices. 

When $n^\mu_\alpha$ and $T^{\mu \nu}_\alpha$ are expressed through 
$n_\alpha$, $\epsilon_\alpha$, $p_\alpha$ and $u^\mu_\alpha$, one faces 
the problem of ordering the matrices. One can define $n^\mu_\alpha$ in 
three ways: $n^\mu_\alpha = n_\alpha u^\mu_\alpha$, 
$n^\mu_\alpha = u^\mu_\alpha n_\alpha$ and 
$n^\mu_\alpha = \{n_\alpha, u^\mu_\alpha \}/2$ where $\{\dots , \dots\}$ 
is the anticommutator. The three different definitions obviously lead to 
three different fluid equations coming from $D_\mu n^\mu_\alpha = 0$. We 
first note that the problem of ordering does not appear when small 
perturbations of colorless state are studied because the linearized 
equations are all the same, as follows from considerations presented 
in Sec.~\ref{sec-res}. The problem of ordering is also absent for 
systems, which are significantly colorful, if the matrices belong to
the Cartan subalgebra of SU($N_c$), as then they commute with each
other. Unfortunately, we have not found a general solution of the 
ordering problem. Numerical simulations will be very helpful to
clarify whether the three formulations are equivalent to each other. 

In the case of an Abelian plasma, the relativistic version of the
Euler equation is obtained from Eq.~(\ref{en-mom-eq}) by removing
from it, following Landau and Lifshitz \cite{Lan63}, the part
which is parallel to $u^\mu_\alpha$. An analogous procedure is not
possible for the non-Abelian plasma because the matrices
$n_\alpha$, $u^\mu_\alpha$, and $u^\nu_\alpha$, in general, do not
commute with each other. Thus, one has to work directly with
Eqs.~(\ref{cont-eq}, \ref{en-mom-eq}) with $n^\mu_\alpha$ and
$T^{\mu \nu}_\alpha$ of the form (\ref{flow-id}, \ref{en-mom-id}).
The equations have to be supplemented by the Yang-Mills equation 
(\ref{yang-mills}) with the color current of the form 
\be
\label{hydro-current} j^\mu(x) = -\frac{g}{2} \sum_\alpha
\Big(n_\alpha u^\mu_\alpha - {1 \over N_c}{\rm Tr}\big[n_\alpha
u^\mu_\alpha  \big]\Big) \;, 
\ee 
where only the quark contribution is taken into account.

The fluid equations (\ref{cont-eq}, \ref{en-mom-eq}) do not form
a closed set of equations even when the chromodynamic field is
treated as an external one. There are five matrix equations
($\nu = 0,1,2,3$) and six unknown matrix functions: $n_\alpha (x)$,
$\epsilon_\alpha (x)$, $p_\alpha (x)$ and three components of
the four-velocity $u_\alpha^\mu(x)$. We note that  the
constraint $u_\alpha^\mu(x)u_{\alpha \,\mu}(x) = 1$ is imposed.

There are several ways to close the system. In the case of real
hydrodynamics, which describes the system in local thermodynamic
equilibrium, one adds the equation of state. Although the system
under consideration is not in equilibrium, we can still add a
relation analogous to the equation of state. The point is that
the energy-momentum tensor (\ref{en-mom}) of a weakly interacting
gas of massless quarks is traceless ($T^\mu_{\alpha \, \mu}(x) = 0$)
because $p^2 = 0$. Then, Eq.~(\ref{en-mom-id}) provides the
desired relation
\be
\label{EoS}
\epsilon_\alpha (x) = 3 p_\alpha (x) \;
\ee
due to the constraint $u_\alpha^\mu(x)u_{\alpha \,\mu}(x) = 1$. 
In fact, the complete energy-momentum tensor of any conformal theory
is traceless but quantum effects, as the trace anomaly, can spoil 
conformal invariance. The phenomenon of mass generation in a medium 
modifies Eq.~(\ref{EoS}) but the effect is small in the perturbative 
regime studied here. We also note that $\epsilon_\alpha (x)$ and 
$p_\alpha (x)$ are {\em not} the energy density and pressure of the 
equilibrium system but the matrices in color space which in equilibrium 
become the energy density and pressure, respectively.

Another method to close the system of equations is to neglect
the gradients of pressure ($p_\alpha (x)$), assuming that the
system's dynamics is dominated by a self-consistently generated 
chromodynamic field. In the following section, where the equations 
are solved in the linear response approximation, we use both methods 
to close the system.

Our analysis is limited to quarks but, as already mentioned,
inclusion of antiquarks and gluons is straightforward. Since the
distribution functions of quarks, antiquarks and gluons of every
stream are assumed to obey the collisionless transport equation,
we have a separated set of fluid equations for quarks, antiquarks
and gluons of every stream. The equations are coupled only through
the chromodynamic mean field. The quarks belong to the fundamental 
representation of the ${\rm SU}(N_c)$ group and thus, the hydrodynamic 
quantities $n_\alpha (x)$, $\epsilon_\alpha (x)$, $p_\alpha (x)$ and 
$u_\alpha^\mu(x)$ are $N_c \times N_c$ matrices. Antiquarks can be 
treated as in  kinetic theory that is as belonging to the fundamental
representation even so they belong, strictly speaking, to the
transposed fundamental representation. Since gluons belong
to the adjoint representation, the hydrodynamic quantities,
and consequently the fluid equations, are
$(N_c^2 -1) \times (N_c^2 -1)$ matrices.


\section{Linear response analysis}
\label{sec-res}


In this section the hydrodynamic equations (\ref{cont-eq}, \ref{en-mom-eq})
are linearized around an stationary, homogeneous and colorless state
described by $\bar n$, $\bar \epsilon$, $\bar p$ and $\bar u^\mu$. We 
mostly skip here the index $\alpha$ to simplify the notation. The index 
is restored in the very final formulas.

Because every stream is assumed to be colorless, the matrices $\bar n$,
$\bar \epsilon$, $\bar p$ and $\bar u^\mu$ are proportional to the
unit matrix in the color space. Thus,
\be
\label{neutral-cov}
\bar n  \, \bar u^\mu
- {1 \over N_c}{\rm Tr}\big[\bar n \, \bar u^\mu  \big]
= 0 \;,
\ee
which means that the color four-current of every stream vanishes.
The quantities of interest are decomposed as
\be
n (x) = \bar n  + \delta n(x)
\;, \;\;\;\;\;\;\;
\epsilon (x) = \bar \epsilon + \delta \epsilon (x)
\;,
\ee
\be
p (x) = \bar p  + \delta p(x)
\;, \;\;\;\;\;\;\;
u^\mu (x) = \bar u^\mu + \delta u^\mu (x)
\;.
\ee
Since the state described by $\bar n$, $\bar \epsilon$, $\bar p$ and
$\bar u^\mu$ is assumed to be stationary, homogeneous and colorless,
we have
\be
D^\mu \bar n  = 0 \;, \;\;\;\;\;\;\;
D^\mu \bar \epsilon  = 0 \;, \;\;\;\;\;\;\;
D^\mu \bar p  = 0 \;, \;\;\;\;\;\;\;
D^\mu \bar u^\nu  = 0 \;. \;\;\;\;\;\;\;
\ee
And because we consider only small deviations from the stationary,
homogeneous and colorless state, the following conditions are obeyed
\be
\bar n \gg \delta n \;, \;\;\;\;\;\;\;
\bar \epsilon \gg \delta \epsilon \;, \;\;\;\;\;\;\;
\bar p \gg \delta p \;, \;\;\;\;\;\;\;
\bar u^\mu \gg \delta u^\mu \;.
\ee
Actually, $\delta n$, $\delta \epsilon$, $\delta p$ and $\delta u^\mu$
should be diagonalized to be comparable to the $\bar n$, $\bar \epsilon$,
$\bar p$ and $\bar u^\mu$.

Substituting the linearized $n^\mu$ and $T^{\mu \nu}$,
which are
\be
\label{flow-lin}
n^\mu  = \bar n \, \bar u^\mu + \bar n \, \delta u^\mu
+ \delta n \,\bar u^\mu
\;,
\ee
\be
\label{en-mom-lin}
T^{\mu \nu} = (\bar \epsilon + \bar p )
\bar u^\mu \, \bar u^\nu - \bar p  \, g^{\mu \nu}
+ (\delta \epsilon + \delta p )
\bar u^\mu \, \bar u^\nu
+ (\bar\epsilon + \bar p )
(\bar u^\mu \, \delta u^\nu + \delta u^\mu \, \bar u^\nu )
- \delta p  \, g^{\mu \nu}
\;,
\ee
into Eqs.~(\ref{cont-eq}, \ref{en-mom-eq}), one finds
\be
\label{cont-lin-eq}
\bar n \, D_\mu \delta u^\mu
+ (D_\mu \delta n ) \, \bar u^\mu = 0 \;,
\ee
\be
\label{en-mom-lin-eq}
\big(D_\mu (\delta \epsilon + \delta p )\big)
\bar u^\mu \, \bar u^\nu
+ (\bar\epsilon + \bar p )
\big(\bar u^\mu \, (D_\mu \delta u^\nu)
+ (D_\mu \delta u^\mu) \, \bar u^\nu \big)
- D^\nu \delta p
- g F^{\mu \nu} \bar n \bar u_\mu = 0 \;.
\ee
Projecting Eq.~(\ref{en-mom-lin-eq}) on $\bar u^\nu$,
one finds
\be
\label{en-den-lin-eq}
\bar u^\mu D_\mu \delta \epsilon
+ (\bar\epsilon + \bar p ) D_\mu \delta u^\mu = 0 \;.
\ee
To derive Eq.~(\ref{en-den-lin-eq}) one has to observe
that $u^\mu u_\mu = 1 = \bar u^\mu  \bar u_\mu$, and
consequently, $\bar u^\mu \delta u_\mu =
{\cal O}\big((\delta u)^2\big)$ and
$\bar u^\mu D^\nu \delta u_\mu
= {\cal O}\big((\delta u)^2\big)$.

Acting on Eq.~(\ref{en-mom-lin-eq}) with the projection
operator $(g_{\sigma \nu} - \bar u_\sigma \bar u_\nu)$,
one gets the linearized relativistic Euler equation
\be
\label{Euler-lin}
(\bar \epsilon + \bar p ) \bar u_\mu D^\mu  \delta u^\nu
- (D^\nu - \bar u^\nu \bar u_\mu D^\mu ) \delta p
- g \bar n  \bar u_\mu F^{\mu \nu}= 0\;.
\ee
As already mentioned,
Eqs.~(\ref{cont-lin-eq},\ref{en-den-lin-eq},\ref{Euler-lin})
do not form a closed set of equations. In the following two sections
we solve the fluid equations
(\ref{cont-lin-eq},\ref{en-den-lin-eq},\ref{Euler-lin}), adopting
two different methods to close the system.

\subsection{Pressure gradients neglected}

The system is closed when the pressure gradients are neglected,
which physically means that the system's dynamics is dominated
by the mean field. When the term with the pressure gradient is
dropped in Eq.~(\ref{Euler-lin}), the equation (\ref{en-den-lin-eq})
effectively decouples from the remaining equations, and one has
to solve only two equations (\ref{cont-lin-eq}, \ref{Euler-lin})
where $\delta \epsilon$ is absent.

In principle, Eqs.~(\ref{cont-lin-eq}, \ref{Euler-lin}) with the
pressure term neglected can be formally solved in a gauge covariant
manner, using the inverse operator of the covariant derivative.
However, we are interested here only in computing the polarization
tensor. Thus, we replace the covariant derivatives by the normal
ones to fully linearize the equations, and the gauge independence
of the result is checked {\it a posteriori}. After performing the
Fourier transformation, which is defined as
\be
f(x) \equiv \int {d^4 k \over (2\pi)^4} \:
e^{-i k\cdot x} f(k) \;,
\ee
Eqs.~(\ref{cont-lin-eq}, \ref{Euler-lin}) get the form
\be
\label{cont-eq3}
\bar u^\mu k_\mu \delta n
+ \bar n k_\mu \delta u^\mu = 0 \;,
\ee
\be
\label{cov-Euler3}
i(\bar \epsilon + \bar p )
\bar u_\mu k^\mu  \delta u^\nu
+ g \bar n  \bar u_\mu F^{\mu \nu}= 0\;.
\ee
Eqs.~(\ref{cont-eq3}, \ref{cov-Euler3}) are easily solved
providing
\be
\delta n = -i g \;
\frac{\bar n^2}{\bar \epsilon + \bar p} \;
\frac{\bar u_\mu k_\nu}{(\bar u \cdot k)^2} \;
F^{\mu \nu} \;,
\ee
\be
\delta u^\nu = i g \;
\frac{\bar n}{\bar \epsilon + \bar p} \;
\frac{\bar u_\mu}{\bar u \cdot k} \;
F^{\mu \nu} \;.
\ee

Since ${\rm Tr}[ F^{\mu \nu}] = 0$, the induced current equals \be
\label{curr-lin} \delta j^\mu = -\frac{g}{2}\sum_\alpha (\bar
n_\alpha \delta u^\mu_\alpha + \delta n_\alpha \, \bar
u^\mu_\alpha )\;,
\ee
where the index labelling the streams
present in the plasma system is restored. Because the linearized
strength tensor equals $F^{\mu \nu}(k) = - ik^\mu A^\nu(k) +
ik^\nu A^\mu(k)$, one finally finds
 \be \delta j^\mu(k) = -
\Pi^{\mu \nu}(k) A_\nu(k) \;, 
\ee 
with
\be 
\label{Pi-hydro}
\Pi^{\mu \nu}(k) = -\frac{g^2}{2} \sum_\alpha \frac{\bar
n_\alpha^2}{\bar \epsilon_\alpha + \bar p_\alpha} \frac{(\bar
u_\alpha \cdot k) (k^\mu \bar u^\nu_\alpha + k^\nu \bar
u^\mu_\alpha ) - k^2  \bar u^\mu_\alpha \bar u^\nu_\alpha - (\bar
u_\alpha \cdot k)^2 g^{\mu \nu} } {(\bar u_\alpha \cdot k)^2} \;.
\ee 
The polarization tensor $\Pi^{\mu \nu}(k)$ is proportional to
the unit matrix in  color space, it is symmetric and transverse
($k_\mu \Pi^{\mu \nu}(k) = 0$), and thus it is gauge independent.

The polarization tensor obtained within the kinetic theory in
analogous approximation \cite{Mrowczynski:2000ed} is
\be
\label{Pi-kinetic}
\Pi^{\mu \nu}(k) = -\frac{g^2}{2} \int_p
f(p) \;
{ (p\cdot k)(k^\mu p^\nu + k^\nu p^\mu) - k^2 p^{\mu} p^{\nu}
- (p\cdot k)^2 g^{\mu\nu} \over(p\cdot k)^2} \;,
\ee
where $f(p)$ is the distribution function of quarks in the
colorless, stationary and homogeneous state.  One observes that
Eq.~(\ref{Pi-kinetic}) transforms into Eq.~(\ref{Pi-hydro}) when
\be
\label{tsunami}
f(p) = \sum_\alpha \bar n_\alpha \, \bar u^0_\alpha \;
\delta^{(3)}\Big({\bf p}
- \frac{\bar \epsilon_\alpha + \bar p_\alpha}
{\bar n_\alpha} \, {\bf u}_\alpha \Big) \;.
\ee

The delta-like distribution function (\ref{tsunami}), and
consequently the approximation, where the pressure gradients
are neglected, hold for systems where the thermal momentum
($p_{\rm thermal}$) is much smaller than the collective momentum
($p_{\rm collec}$) of the hydrodynamic flow. In the electromagnetic
plasma of electrons and ions, such a situation occurs for a
sufficiently low temperatures when the effects of pressure
are indeed expected to be small. For the system of massless
partons, the condition $p_{\rm thermal} \ll p_{\rm collec}$
is achieved by requiring that $p_{\rm collec}$ is large
rather than $p_{\rm thermal}$ is small. For massless particles
in local equilibrium $p_{\rm thermal} \sim T$, where $T$ is the
local temperature, while the formula (\ref{tsunami}) clearly
shows that $p_{\rm collec} \sim T \bar\gamma_\alpha \bar v_\alpha$
where $\bar v_\alpha $ and $\bar\gamma_\alpha$ are the velocity
and Lorentz factor of the collective flow. Therefore, the condition
$p_{\rm collec} \gg p_{\rm thermal}$ requires
$\bar\gamma_\alpha \gg 1$.

\subsection{Effect of pressure gradients included}

The equation (\ref{EoS}) allows one to close the system of fluid
equations, not neglecting the pressure gradients. Using the
relation (\ref{EoS}), one has to solve three equations
(\ref{cont-lin-eq},\ref{en-den-lin-eq},\ref{Euler-lin}).

Performing the linearization analogous to that from the previous
section and the Fourier transformation, the equations
(\ref{cont-lin-eq},\ref{en-den-lin-eq},\ref{Euler-lin})
get the form
\be
\label{cont-eq6}
\bar u^\mu k_\mu \delta n
+ \bar n k_\mu \delta u^\mu = 0 \;,
\ee
\be
\label{ener-den-eq6}
\bar u^\mu k_\mu \delta \epsilon +
\frac{4}{3}\bar \epsilon k_\mu  \delta u^\mu
= 0\;,
\ee
\be
\label{cov-Euler6}
i\frac{4}{3}\bar\epsilon
\bar u_\mu k^\mu  \delta u^\nu
+ i\frac{1}{3}
(\bar u^\mu \bar u^\nu k_\mu - k^\nu )
\delta \epsilon
+ g\bar n \bar u_\mu F^{\mu \nu}= 0\;.
\ee

Substituting $\delta \epsilon$ obtained from Eq.~(\ref{ener-den-eq6})
into the Euler equation (\ref{cov-Euler6}), one gets
\be
\label{cov-Euler7}
\Big[g^{\nu \mu} + \frac{1}{3(\bar u \cdot k)^2}
\big(k^\nu k^\mu - \bar u^\nu k^\mu (\bar u \cdot k) \big) \Big]
\delta u_\mu =
i\frac{3}{4} g \,
\frac{\bar n}{\bar\epsilon \,(\bar u \cdot k)} \,
\bar u_\mu F^{\mu \nu} \;.
\ee
Observing that
$$
\big(k_\sigma k_\nu - \bar u_\sigma k_\nu(\bar u \cdot k) \big)
\big(k^\nu k_\mu - \bar u^\nu k_\mu (\bar u \cdot k) \big)
\propto (k_\sigma k_\mu - \bar u_\sigma k_\mu(\bar u \cdot k)\big) \;,
$$
the operator in the left-hand-side of Eq.~(\ref{cov-Euler7}) can
be inverted as
\be
\Big[g_{\sigma \nu} - \frac{1}{k^2 + 2(\bar u \cdot k)^2}
\big(k_\sigma k_\nu - \bar u_\sigma k_\nu
(\bar u \cdot k) \big) \Big] \;
\Big[g^{\nu \mu} + \frac{1}{3(\bar u \cdot k)^2}
\big(k^\nu k^\mu - \bar u^\nu k^\mu (\bar u \cdot k) \big) \Big]
= g_\sigma^{\;\;\mu} \;,
\ee
and Eq.~(\ref{cov-Euler7}) is exactly solved by
\be
\delta u_\sigma = i\frac{3}{4} g \,
\frac{\bar n}{\bar\epsilon \,(\bar u \cdot k)}
\Big[g_{\sigma \nu} - \frac{1}{k^2 + 2(\bar u \cdot k)^2}
\big(k_\sigma k_\nu - \bar u_\sigma k_\nu
(\bar u \cdot k) \big) \Big]
\bar u_\mu F^{\mu \nu} \;.
\ee

Substituting $\delta n$ and $\delta u^\mu$ into the current
(\ref{curr-lin}), one finds the polarization tensor as
\ba
\label{Pi2}
\nonumber
\Pi^{\mu \nu}(k) = -\frac{g^2}{2} \sum_\alpha
\frac{3 \bar n_\alpha^2}{4 \bar\epsilon_\alpha}
\frac{1}{(\bar u_\alpha \cdot k)^2}
&\Big[&
(\bar u_\alpha \cdot k)
(k^\mu \bar u^\nu_\alpha + k^\nu \bar u^\mu_\alpha )
- (\bar u_\alpha \cdot k)^2 g^{\mu \nu}
- k^2  \bar u^\mu_\alpha \bar u^\nu_\alpha \\[2mm]
&-&
\frac{(\bar u_\alpha \cdot k)k^2
(k^\mu \bar u^\nu_\alpha + k^\nu \bar u^\mu_\alpha )
- (\bar u_\alpha \cdot k)^2 k^\mu k^\nu
- k^4  \bar u^\mu_\alpha \bar u^\nu_\alpha}
{k^2 + 2(\bar u_\alpha \cdot k)^2} \: \Big]\;,
\ea
where the stream index $\alpha$ has been restored. The first
term in Eq.~(\ref{Pi2}) corresponds to the polarization
tensor (\ref{Pi-hydro}) found when the pressure gradients are
neglected while the second term gives the effect of the pressure
gradients. The first term is, as already mentioned, symmetric
and transverse. The second term is symmetric and transverse as
well. Thus, the whole polarization tensor (\ref{Pi2}) is symmetric
and transverse.

As explained below Eq.~(\ref{tsunami}), the pressure gradients can
be neglected and the distribution function can be approximated by
the delta-like form (\ref{tsunami}) when $\bar \gamma_\alpha \gg 1$.
Because the four-velocity is expressed as $ \bar u^\mu_\alpha =
(\bar\gamma_\alpha ,\bar\gamma_\alpha \bar{\bf  v}_\alpha)$, it
is easy to check that the second term in Eq.~(\ref{Pi2}) is small,
when compared to the first one, for $\bar \gamma_\alpha \gg 1$.
However, one has to be careful here: when the wave vector ${\bf k}$
is exactly parallel to $\bar{\bf v}_\alpha$ and both are, say, in
the $z$ direction, then the two terms of the non-zero components of
the polarization tensor (\ref{Pi2}) ($\Pi^{00}$, $\Pi^{0z}$,
$\Pi^{zz}$) vanish as $1/\bar\gamma_\alpha^2$. For the remaining
non-zero components of (\ref{Pi2}) ($\Pi^{xx}$, $\Pi^{yy}$),
the second term is identically zero. We return to this point when
the collective modes are discussed.


\section{Collective modes in the two-stream system}
\label{sec-2stream}


As an application of the polarization tensors derived in the previous
section, we discuss here collective plasma modes in the two-stream
system. Within  kinetic theory such a system was studied in 
\cite{Pokrovsky:1988bm,Pokrovsky:1990sz,Pokrovsky:1990uh,Mrowczynski:1988dz}.
The two-stream configuration has not much to do with the experimental 
situation in heavy-ion collisions where the momentum distribution of 
the produced partons is very different from the two-peak shape 
characteristic for the two-stream system. Thus, it should be treated only 
as a toy model. However, the model dynamics is rather nontrivial due to
several unstable modes. While the two-stream configuration is rather of
academic interest, one can model experimentally relevant situations with 
several streams. And in the linear regime, such a multi-stream configuration 
can be studied along the lines presented below. 

Since Fourier transformed chromodynamic field $A^{\mu}(k)$ satisfies the
equation of motion as
\be
\label{eq-A}
\Big[ k^2 g^{\mu \nu} -k^{\mu} k^{\nu} - \Pi^{\mu \nu}(k) \Big]
A_{\nu}(k) = 0 \;,
\ee
the general dispersion equation of the collective plasma modes is
\be
\label{dispersion-pi}
{\rm det}\Big[ k^2 g^{\mu \nu} -k^{\mu} k^{\nu} - \Pi^{\mu \nu}(k) \Big]
 = 0 \;.
\ee
Due to the transversality of $\Pi^{\mu \nu}(k)$ not all components of
$\Pi^{\mu \nu}(k)$ are independent from each other, and consequently
the dispersion equation (\ref{dispersion-pi}), which involves a determinant
of a $4\times4$ matrix, can be simplified to the determinant of a
$3\times3$ matrix. For this purpose one usually introduces the color
dielectric tensor $\varepsilon^{ij}(k)$, where the indices
$i,j = 1,2,3$ label three-vector and tensor components, which is
expressed through the polarization tensor as
\ban
\epsilon^{ij}(k) = \delta^{ij} + {1 \over \omega^2} \Pi^{ij}(k) \;,
\ean
where $k \equiv (\omega, {\bf k})$. Then, the dispersion equation
gets the form
\be
\label{dispersion-g}
{\rm det}\Big[ {\bf k}^2 \delta^{ij} -k^i  k^j
- \omega^2 \varepsilon^{ij}(k)  \Big]  = 0 \,.
\ee
The relationship between Eq.~(\ref{dispersion-pi}) and
Eq.~(\ref{dispersion-g}) is most easily seen in the Coulomb
gauge when $A^0 = 0$ and ${\bf k} \cdot {\bf A}(k)=0$. Then,
${\bf E} = i\omega {\bf A}$ and Eq.~(\ref{eq-A}) is immediately
transformed into an equation of motion of ${\bf E}(k)$ which
further provides the dispersion equation (\ref{dispersion-g}).

The dielectric tensor given by the polarization tensor
(\ref{Pi-hydro}), which neglects the effect of pressure, is
\be
\label{epsilon-hydro}
\varepsilon^{ij}(\omega,{\bf k}) =
\Big(1 - \frac{\omega_p^2}{\omega^2} \Big) \delta^{ij}
- \frac{g^2}{2\omega^2}
\sum_\alpha
\frac{\bar n_\alpha^2}{\bar \epsilon_\alpha + \bar p_\alpha}
\bigg( \frac{\bar v_\alpha^i k^j +\bar v_\alpha^j k^i}
{\omega - {\bf k} \cdot \bar {\bf v}_\alpha}
- \frac{(\omega^2 - {\bf k}^2) \bar v_\alpha^i \bar v_\alpha^j}
{(\omega - {\bf k} \cdot \bar {\bf v}_\alpha )^2} \bigg) \;,
\ee
where $\bar {\bf v}_\alpha$ is the hydrodynamic velocity related
to the hydrodynamic four-velocity $\bar u^\mu_\alpha$; $\omega_p$ 
is the plasma frequency given as
\be
\omega_p^2 \equiv \frac{g^2}{2}\sum_\alpha
\frac{\bar n_\alpha^2}{\bar \epsilon_\alpha + \bar p_\alpha} \;.
\ee

The dielectric tensor given by the polarization tensor (\ref{Pi2}),
which includes the effect of pressure, is
\ba
\label{epsilon-hydro2}
\varepsilon^{ij}(\omega,{\bf k}) =
\Big(1 - \frac{\omega_p^2}{\omega^2} \Big) \delta^{ij}
&-& \frac{3g^2}{8\omega^2}
\sum_\alpha
\frac{\bar n_\alpha^2}{\bar \epsilon_\alpha}
\bigg( \frac{\bar v_\alpha^i k^j +\bar v_\alpha^j k^i}
{\omega - {\bf k} \cdot \bar {\bf v}_\alpha}
- \frac{(\omega^2 - {\bf k}^2)\bar v_\alpha^i \bar v_\alpha^j}
{(\omega - {\bf k} \cdot \bar {\bf v}_\alpha )^2}
\\ \nonumber
&-& \frac{(\omega - {\bf k} \cdot \bar {\bf v}_\alpha)
(\omega^2 - {\bf k}^2)(\bar v_\alpha^i k^j +\bar v_\alpha^j k^i)
- (\omega - {\bf k} \cdot \bar {\bf v}_\alpha)^2k^i k^j
-(\omega^2 - {\bf k}^2)^2\bar v_\alpha^i \bar v_\alpha^j}
{(\omega - {\bf k} \cdot \bar {\bf v}_\alpha)^2\big(\omega^2 - {\bf k}^2
+ 2 \bar\gamma_\alpha^2 (\omega - {\bf k} \cdot \bar {\bf v}_\alpha)^2\big)}
\bigg) \;,
\ea
with the plasma frequency given as
\be
\omega_p^2 \equiv \frac{3g^2}{8}\sum_\alpha
\frac{\bar n_\alpha^2}{\bar \epsilon_\alpha} \;.
\ee

In the following subsections we consider the collective modes in the
two-stream system with the wave vectors perpendicular and parallel to
the hydrodynamic velocity. The velocities $\bar {\bf v}_\alpha $ are
chosen to be oriented along the axis $z$, {\it i.e.}
$\bar {\bf v}_\alpha = (0,0,\bar v_\alpha)$. The index $\alpha$, which
labels the streams has two values, $\alpha = \pm$. For simplicity we
also assume that
\be
\label{streams}
\bar v \equiv \bar v_+ = -\bar v_- \;, \;\;\;\;\;\;\;
\bar n \equiv \bar n_+ = \bar n_- \;, \;\;\;\;\;\;\;
\bar \epsilon \equiv \bar \epsilon_+ = \bar \epsilon_- \;, \;\;\;\;\;\;\;
\bar p \equiv \bar p_+ = \bar p_- \;.
\ee
Then, the plasma frequency equals
$\omega_p^2 = g^2 \bar n^2 /(\bar \epsilon + \bar p)$ or
$\omega_p^2 = 3g^2 \bar n^2 /(4\bar \epsilon)$.

\subsection{$\;\;\;{\bf k} \perp \bar{\bf v}$}

The wave vector is chosen to be parallel to the axis $x$,
${\bf k} = (k,0,0)$. Due the conditions (\ref{streams}),
the off-diagonal elements of the matrix in Eq.~(\ref{dispersion-g})
vanish.

\subsubsection{No pressure gradient effect}

With the dielectric tensor given by Eq.~(\ref{epsilon-hydro}),
the dispersion equation is
\be
(\omega^2 - \omega_p^2)(\omega^2 - \omega_p^2 - k^2)
\Big(\omega^2 - \omega_p^2 - k^2
- \lambda^2  \; \frac{k^2 - \omega^2}{\omega^2} \Big) = 0 \;,
\ee
where $\lambda^2 \equiv \omega_p^2 \bar v^2$. As solutions of the
equation, one finds the stable longitudinal mode with
$\omega^2 = \omega_p^2$ and the stable transverse mode with
$\omega^2 = \omega_p^2 + k^2$. There are also transverse modes with
\be
\label{solut11}
\omega_{\pm}^2 = \frac{1}{2}\Big(
\omega_p^2 - \lambda^2 + k^2 \pm
\sqrt{(\omega_p^2 - \lambda^2 + k^2)^2 + 4 \lambda^2 k^2} \, \Big)
\;.
\ee
As seen, $\omega_+^2 > 0$ but $\omega_-^2 < 0$. Thus, the mode $\omega_+$
is stable and there are two modes with pure imaginary frequency corresponding
to $\omega_-^2 < 0$. The first mode is overdamped while the second one
is the well-known unstable Weibel mode leading to the filamentation
instability.

As discussed below Eq.~(\ref{tsunami}), the pressure gradients can be
neglected when $\bar \gamma \gg 1$, and thus, the solutions (\ref{solut11})
are physically relevant in this limit. Therefore, we write down the
solutions (\ref{solut11}) for $\bar \gamma \gg 1$ which are
\be
\label{solut12}
\omega_{\pm}^2 = \frac{1}{2}\Big(
k^2 \pm \sqrt{k^4 + 4 \omega_p^2 k^2} \, \Big)
\;.
\ee
As seen in Eq.~(\ref{solut12}), the modes are independent of $\bar \gamma$,
if it is sufficiently large.

\subsubsection{Effect of pressure gradients included}

As in the `pressureless' case, there are two stable modes
$\omega^2 = \omega_p^2$ and the stable transverse mode with
$\omega^2 = \omega_p^2 + k^2$. The transverse modes corresponding to
those given by Eq.~(\ref{solut11}) are obtained by solving the equation
\be
\omega^4 - \omega^2 (k^2 + \omega^2_p )
+ \lambda^2 (\omega^2 - k^2) -
\lambda^2 \frac{(\omega^2 - k^2)^2}{\omega^2 - k^2
+ 2 \bar \gamma^2 \omega^2} = 0 \;.
\ee
Defining the dimensionless variables $a \equiv \omega/\omega_p$ and
$b \equiv k/\omega_p$, the above equation can then be rewritten as
follows
\be
\label{solut33}
a^2 \big[ a^4 - F(b,\bar v) a^2 -  G(b,\bar v) \big] = 0 \;,
\ee
where
\be
F(b,\bar v) \equiv \frac{ (3 + 2 b^2)(1-\bar v^2) + 2b^2}{3 - \bar v^2} \ ,
\qquad G(b,\bar v) \equiv b^2\frac{ 3\bar v^2 -1- b^2(1- \bar v^2)}
{3 - \bar v^2} \;.
\ee
Solutions of Eq.~(\ref{solut33}) are given by $a^2 =0$ and
\be
\label{solut22}
a^2_\pm = \frac{1}{2} \big(F \pm \sqrt{F^2 + 4G}\,\big)
\ee
The solution with $a^2_-$ describes the unstable mode
($a^2_- < 0$) provided $G$ is positive. The condition
$G > 0$ gives
\be
\bar v^2 > \frac{1}{3}
\;\;\;\;\;\; {\rm and} \;\;\;\;\;\;
k^2 < \frac{3 \bar v^2 - 1}{1 -\bar v^2} \, \omega_p^2 \;.
\ee
As seen, the instability appears when the stream velocity
is larger than the speed of sound in the ideal gas of massless
partons and when the wavelength is sufficiently long. We also
note that in the limit $\bar \gamma \gg 1$ the solutions
(\ref{solut22}) reproduce, as expected, those given by
Eq.~(\ref{solut12}).

\subsection{$\;\;\;\;{\bf k} \parallel \bar{\bf v}$}

The velocities and the wave vector are chosen to be oriented along
the $z-$axis {\it i.e.} $\bar {\bf v}_\pm = (0,0,\pm \bar v)$
and ${\bf k} = (0,0,k)$. Then, the matrix in Eq.~ (\ref{dispersion-g})
is diagonal. As in the case ${\bf k} \perp \bar{\bf v}$,
we discuss separately the collective modes given by the dielectric
tensors (\ref{epsilon-hydro}) and (\ref{epsilon-hydro2}). However,
it is physically rather unjustified because, as explained at the
very end of Sec.~\ref{sec-res}, the tensor (\ref{epsilon-hydro})
cannot be treated as an approximation of (\ref{epsilon-hydro2})
even for $\gamma \gg 1$. The effects of the mean-field and of the
pressure gradients appear to be equally important. Therefore,
only the results of subsection \ref{sec777} are physically
reliable.

\subsubsection{No pressure gradient effect}

With the dielectric tensor given by Eq.~(\ref{epsilon-hydro}),
the dispersion equation reads
\be
\label{disper-eq22}
(\omega^2 - \omega_p^2 - k^2)^2\bigg(
\omega^2 - \omega_p^2
- \omega_p^2 \Big(\frac{ k \bar v} {\omega - k \bar v} +
\frac{(k^2 - \omega^2)\bar v^2} {2(\omega - k \bar v)^2}
- \frac{ k\bar v}{\omega + k \bar v} +
\frac{(k^2 - \omega^2)\bar v^2} {2(\omega + k \bar v )^2} \Big)
\bigg)= 0 \;.
\ee
There are two transverse stable modes with $\omega^2 = \omega_p^2 + k^2$.
The longitudinal modes are solutions of the equation which can be
rewritten as
\be
\label{disper-2stream}
1 - \omega_0^2\Big(
\frac{1}{(\omega - k \bar v)^2}
+\frac{1}{(\omega + k \bar v )^2} \Big) = 0 \;,
\ee
where $\omega_0^2 \equiv \omega_p^2/2\bar \gamma^2$.
With the dimensionless quantities $x\equiv \omega/\omega_0$,
$y \equiv k \bar v/ \omega_0$, Eq.~(\ref{disper-2stream}) is
\be
(x^2 - y^2)^2 - 2x^2 - 2y^2 = 0 \;,
\ee
and it is solved by
\be
x_\pm^2 = y^2 + 1 \pm \sqrt{4y^2 + 1} \;.
\ee
As seen, $x_+^2$ is always positive and thus, it gives two
real (stable) modes; $x_-^2$ is negative for $ 0 < y < \sqrt{2}$
and then, there are two pure imaginary modes. The unstable one
corresponds to the two-stream electrostatic instability.

\subsubsection{Effect of pressure gradients included}
\label{sec777}

The transverse modes are the same as in the `pressureless' case while
the longitudinal modes are solutions of the equation
\ba
\nonumber
1 & - & \omega^2_0\bigg(
\frac{1}{(\omega - k \bar v)^2}
+ \frac{1}{(\omega + k \bar v)^2} \bigg) \\[2mm]
& + & \frac{\omega^2_p}{2}
\bigg( \frac{(\omega \bar v - k)^2}
{(\omega - k \bar v)^2 (2 \bar \gamma^2 (\omega - k \bar v)^2
+ \omega^2 - k^2)}
+ \frac{(\omega \bar v + k)^2}{(\omega + k \bar v)^2
(2 \bar \gamma^2 (\omega + k \bar v)^2 + \omega^2 - k^2)} \bigg)
= 0 \;,
\ea
which, in terms of the dimensionless quantities introduced in the
previous section, is
\be
1 - \frac{1}{(x -y)^2}
\bigg( 1 + \frac{(\bar v x -y/\bar v)^2}{2 (x -y)^2 +
\frac{1}{\bar \gamma^2 } (x^2 - y^2/\bar v^2 )}\bigg)
- \frac{1}{(x +y)^2}
\bigg( 1 + \frac{(\bar v x +y/\bar v)^2}{2 (x +y)^2 +
\frac{1}{\bar \gamma^2 } (x^2 -  y^2/\bar v^2)} \bigg) = 0 \;.
\ee
The equation can be converted to
\be
(x^2 -y^2)^2 \big[x^4 - B(y,\bar v) x^2 - C(y, \bar v) \big] = 0 \,
\ee
with
\be
B(y, \bar v) \equiv \frac{18 + 6 \bar v^2 y^2 + 6y^2 /\bar v^2
-4 y^2 -6 \bar v^2}{(3 - \bar v^2)^2} \ , \qquad
C(y, \bar v) \equiv  y^2\frac{ \left(3 - \frac{1}{\bar v^2}\right)
[ 6 - y^2 \left(3 - \frac{1}{\bar v^2}\right)]
}{(3 -\bar v^2)^2}
\;,
\ee
which is solved by $x^2 = y^2$ and
\be
x^2_\pm = \frac{1}{2}\big(B \pm \sqrt{B^2 + 4C}\,\big) \;.
\ee
As for the case when the wave vector is perpendicular to the velocity,
there are only instabilities when the stream velocity is bigger than the
speed of sound in the ideal gas of massless partons, as $C > 0$ is satisfied 
when
\be
\bar v^2 > \frac{1}{3}
\;\;\;\;\;\; {\rm and} \;\;\;\;\;\;
k^2 < 3 \frac{1 -\bar v^2}{3 \bar v^2 - 1} \, \omega_p^2 \;.
\ee
It is also interesting to note that for the particular case $\bar v^2=1/3$,
$C=0$, and the solutions of the dispersion relations get a very simple
form. Namely, we have $\omega^2_- = 0$ and
\be
\omega^2_+ = \frac 34 \left( \omega_p^2 + k^2 \right) \;.
\ee


\section{Concluding remarks}


The fluid equations (\ref{cont-eq},\ref{en-mom-eq}) with $n^\mu_\alpha$
and $T^{\mu \nu}_\alpha$ given by the formulas (\ref{flow-id},\ref{en-mom-id})
form a closed set of equations when supplemented by the relation (\ref{EoS}).
If the chromodynamic field is generated self-consistently by the color fluid,
the Yang-Mills equation (\ref{yang-mills}) with the color current of the
form (\ref{hydro-current}) must be added. The derivation of these equations
is the main result of our paper. Although some information is lost when
going from  kinetic theory to hydrodynamics, the dynamical content of the 
fluid equations is still very rich. The results of Sec.~\ref{sec-2stream},
where the two-stream system is studied, demonstrate it convincingly.
The system is shown to be unstable with respect to the magnetic and 
electric modes, if the hydrodynamic velocity is sufficiently large.

As noted in the Introduction, the hydrodynamic approach is frequently used
in studies of electromagnetic plasmas. Numerical simulations of the fluid
equations are much simpler than those of  kinetic theory because the
distribution function depends on four-position and momentum while in the
fluid approach one deals with a few functions of four-position only. For
this reason we expect that the equations derived here will be useful in
numerical studies of the unstable quark-gluon plasma. Hopefully, the late
stage of instability development, when the dynamics is strongly nonlinear,
can be attacked within the fluid approach.  In the case of the two-stream
system, our analytic results from Sec.~\ref{sec-2stream} can be used to gauge
numerical calculations. While the two-stream configuration is rather academic, one can model situations, which are relevant for quark-gluon plasma from relativistic heavy-ion collisions, with several streams. And then, our fluid 
equations can be applied. 

We also note that the hydrodynamic approach presented here can 
be extended to include the effect of collisions which can be modeled with
transport coefficients (viscosities, conductivities) and additional
terms in the fluid equations. Then, the approach would be applicable to
longer time scales. 

Our derivation of the chromo-hydrodynamic equations is based on the
parton kinetic theory valid in the regime of very weak couplings or, 
equivalently, of very high temperatures $T \gg T_c$, where $T_c$ is 
the temperature of the deconfinement phase transition. One may 
wonder what happens in the regime of not so weak couplings. Because 
the hydrodynamic equations are the expressions of the conservation 
laws, one can hope that the structure of the equations survives when 
the plasma is no longer weakly coupled. Lattice studies indicate that 
an equilibrium quark-gluon plasma at temperatures close to $T_c$ is 
very different from the ideal gas of massless quasiparticles 
\cite{Karsch:2001vs}. The deviations can be, at least partially, 
accounted for in a quasiparticle model when the gluon dispersion 
relation is modified, see e.g.
\cite{Gorenstein:1995vm,Zwanziger:2004np,Bluhm:2004xn}. In such a case, 
our approach can be easily adapted for by changing the relation (\ref{EoS}). 
We hope to report about those results in the near future.

\section{Acknowledgments}

We are grateful to Massimo Mannarelli and Mike Strickland for fruitful 
discussions. C.M. was supported by MEC under grants FPA2004-00996 and 
AYA 2005-08013-C03-02, and by GVA under grant GV05/164.


\end{document}